# Photoinduced modulation of refractive index in Langmuir-Blodgett films of Azo-based H-shaped liquid crystal molecules

Ashutosh Joshi[1], Akash Gayakwad[1], Manjuladevi V.[1], Mahesh C. Varia[2], S. Kumar[2,3], R. K. Gupta[1,*]

[1]*Department of Physics, Birla Institute of Technology and Science, Pilani (BITS-Pilani), Rajasthan, India, 333031*

[2]*Raman Research Institute, Sadashivanagar, Bangalore – 560080*

[3]*Deparment of Chemistry, Nitte Meenakshi Institute of Technology, Bangalore - 560064*

*Corresponding Author Contact: \*raj@pilani.bits-pilani.ac.in*

**Abstract**

The development of optically active area consisting of organic molecules are essential for the devices like optical switches and waveguides, as it can be easily maneuvered by the application of suitable electromagnetic (EM) waves. In this article, we report the development of a photoactive surface by the deposition of a single layer of Langmuir-Blodgett (LB) film of a novel H-shaped liquid crystal (HLC) molecule. The synthesized HLC molecules possess azo-groups and nitro-groups. The azo-group can be isomerized (trans-cis transformation) by irradiating them with ultraviolet (UV) light. The nitro-group can provide sufficient amphiphilicity to the HLC molecules to form a stable Langmuir monolayer at air-water interface. The Langmuir monolayer of the HLC molecules exhibited gas and liquid-like phases. A single layer of LB film of HLC molecules was deposited on a gold chip of a home-built surface plasmon resonance (SPR) instrument. The azo-groups of the molecules in LB film was excited by UV irradiation leading to a change in morphology due to trans-cis transformation. Such a change in morphology can lead to a miniscule change in refractive index (RI) of the LB film. SPR is a label free and highly sensitive optical phenomenon for the measurement of such changes in RI. In our studies, we found systematic changes in the resonance angle of the LB film of HLC molecules as a function of intensity of the UV irradiation. We measured switch-on and switch-off intensity which may suggest that the LB film of HLC molecules can find applications in optical switches or waveguides.

## 1. Introduction

The control of physicochemical properties of materials due to external parameters viz. electric, magnetic fields or electromagnetic (EM) waves are essential for the design and development of novel devices. One of the popular mechanisms is to obtain the control by designing the photoactive organic molecules. Mostly, chromophore is chemically attached to the molecules which can be excited by the absorption of the suitable EM wave. The molecule can de-excite to a lower energy level by a radiative process with the emission of an EM wave of lower energy or non-radiative process by changing the morphology of the molecules. Azo-group (-N=N-) is one of the groups which can be a morphological transformation from trans-to-cis by the exciting them using EM wave of wavelength in ultraviolet range [1–8]. The molecule can relax back to trans conformation either by exposure to ambient light or even in the dark state [9]. The morphological transformation of organic molecules promises several unique device applications viz. optical switches, molecular motors, waveguides in photonics etc [6,10–14]. The mesophases of liquid crystal (LC) molecules can be influenced by the external parameters like temperature, pressure, electric and magnetic fields. The structure-property relationship of the liquid crystal molecules is important to obtain LC based high performing devices. The chemical structures of the molecules can be changed to introduce new phases and enhance liquid crystalline properties in a given mesophase. The liquid crystals molecules possessing azo-group are very interesting as morphology of the molecules can be changed under the influence of EM waves leading to changes in the bulk liquid crystalline phases. Two-dimensionally confined monolayers are good candidates for the development of next-generation flexible and transparent optoelectronics devices [15–18]. Due to enhanced physicochemical properties of the ultrathin films, its activity is extremely high as compared to the thick layers or bulk materials. It is therefore interesting to study the surface behavior of organic molecules exhibiting functional groups which can be tapped through external parameters to deliver high performing devices. The deposition of ultrathin film at air-solid interfaces using Langmuir-Blodgett (LB) methodology are very interesting as they possess a huge potential for industrial application. It can be applied as both active and passive layer for a number device fabrication. In the field of sensors, it has been reported that organized and ultrathin nature of LB films of materials can offer a large enhancement in sensing parameters as compared to randomly oriented thick films [19,20]. The LB films can be used for non-linear optical devices, photovoltaics, ultrafiltration membrane, molecular electronic, and energy storage devices [21–23]. The shape-anisotropic liquid crystal molecules at the air-water interface show very interesting results. A single layer of LC molecules at an interface offers a highly in-plane anisotropy in optical properties [24]. The low in-plane elastic modulus of the film of LC molecules facilitates easy geometrical perturbation by the application of external electric, magnetic fields and EM waves [25]. Such geometrical perturbation can influence the optical and electrical properties and thereby provides an avenue for controlling the physical properties of the film by such external parameters.

There are several forms of shape anisotropic LC molecules which can form a stable Langmuir monolayer at the air-water interface and show a variety of interesting phenomena. There are several studies on Langmuir monolayer and LB films of rod shaped [26], disc-shaped [27], and bow-shaped [25,28] liquid crystal molecules. There are some studies on non-traditional LC molecules [29–31]. In this article, we report our studies on H-shaped LC molecules consisting of azo-groups and several chiral centers. The presence of two nitro-groups provides sufficient amphiphilicity to the molecules to form a stable Langmuir monolayer at the air-water interface. The monolayer in the liquid-like phase on the water subphase is transferred to solid substrate by the highly controlled Langmuir-Blodgett (LB) technique [32]. The morphology of the LB film was obtained using field emission scanning electron microscope (FESEM). The azo-groups of the HLC were excited by irradiating the LB film using a UV source. This facilitates the trans-to-cis transformation. Such morphological transformation can lead to change in refractive index (RI) of the LB film which was studied using a very high-resolution surface plasmon resonance instrument. Here, we report systematic changes in the RI by changing the intensity of the UV irradiation. We found a cut-off and saturation intensity which might be an indicator for the development of optical switches or RI modulated photonic waveguides. In our earlier study, we reported that a polarization of an incident EM wave induced a change in morphology in a layered structure which can be studied using SPR phenomenon [33]. SPR is a label free highly sensitive optical phenomenon which can yield a perceptible change in RI due to miniscule changes in RI of the film deposited onto the gold chip of SPR instrument [34].

2. Experimental

The H-shaped liquid crystal (HLC) molecule was synthesized in the laboratory. The synthesis is briefly described in the supplementary information. The HLC molecule exhibits liquid crystalline phases as : crystal 45 ºC smectic A 78 ºC isotropic. The molecule exhibited nitro groups which can provide sufficient amphiphilicity for them to form a stable Langmuir monolayer at the air-water interface. The chemical structure of the molecule is shown in Fig. 1.

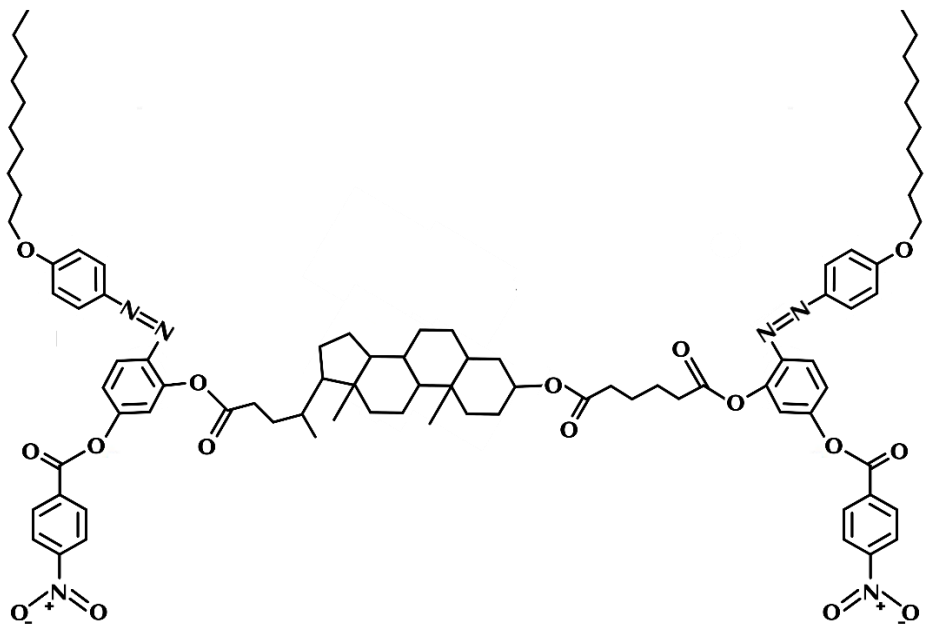

**Fig. 1.** Chemical structure of H-shaped mesogenic liquid crystal molecule (HLC) Bis [5 – (4 – n – dodecyloxybenzoyloxy) – 2 – (4 – methylphenylazo) phenyl] adipate.

A 0.5 mg/ml clear solution was obtained by dissolving the HLC molecules in the high-performance liquid chromatography (HPLC) grade chloroform (Merck). The molecules were spread onto quartz substrate by drop-casting method and UV absorption spectrum of the molecules was recorded in the transmission mode. The solution of the HLC molecules was spread dropwise using a microsyringe (Hamilton) on the surface of ultrapure ion-free water (MilliQ) in a Langmuir-Blodgett trough (KSV NIMA). The trough was equipped with coupled double barriers for symmetric compression of the monolayer. About 15 minutes time was allowed for the solvent to evaporate from the surface of water leaving behind the dispersed HLC molecules. The monolayer was compressed at a speed of 5 mm/min and the surface pressure ($\pi$) -area per molecule ($A_m$) isotherm was recorded. The monolayer at the air-water interface was imaged using a Brewster angle microscope (MicroBAM, KSV NIMA). The Brewster angle microscope (BAM) was equipped with a 50 mW laser of wavelength 659 nm. The Langmuir monolayer of the HLC molecules was transferred onto solid substrates through the LB technique at a target surface pressure ($\pi_t$) of 15 mN/m. The substrates used in our studies were one side polished silicon wafers (Ted Pella), gold deposited quartz wafers (SRS, USA) and the SPR chip consisting of Cr/Au layers deposited on BK7 glass (RI=1.51) substrates. The SPR chips were fabricated in the laboratory by depositing chromium (Cr) film of thickness 4 nm followed by a gold layer of thickness 50 nm on the BK7 glass substrates. The metal deposition was done using a DC sputter (Quorum). The morphology of the LB films of the HLC molecules was obtained using a field emission scanning electron microscope (FESEM, Zeiss Sigma). The SPR measurements were performed using a home-built setup developed in Kretschmann configuration [35]. This configuration is based on angular interrogation wherein angle of incidence is changed at a high resolution and reflected intensity recorded simultaneously. The angular resolution of the setup was 5 µrad. Using the standard glucose solutions, the sensitivity of the equipment was obtained as ~195 °/RIU.

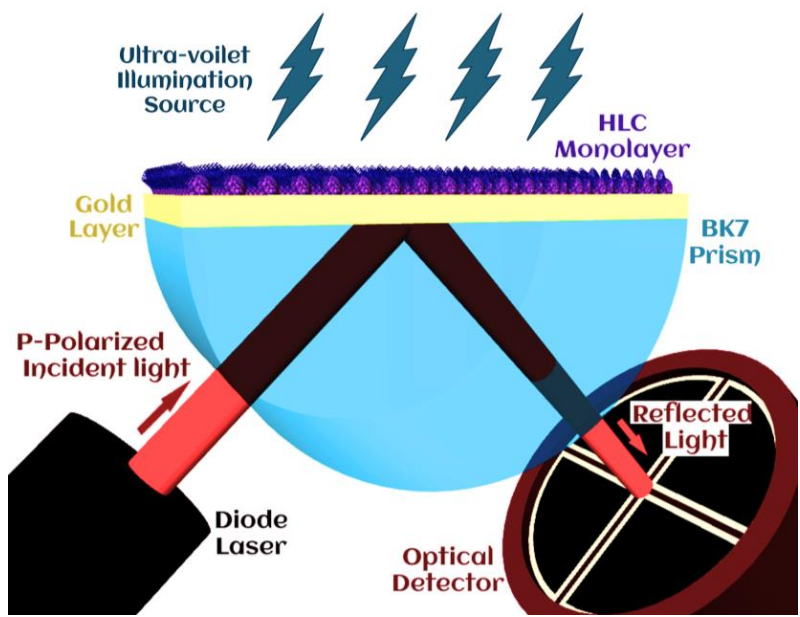

**Fig. 2.** Upgraded SPR setup with UV illumination

In order to irradiate the LB film of the HLC molecules with UV source during the SPR measurement, the SPR instrument was modified as shown in the schematic (Fig. 2). In the SPR setup, a provision was made to irradiate the LB film deposited onto the gold chip using an external UV source. The UV source exhibited a spectrum (Fig. S2 in SI). The spectrum reveals a predominant peak at around 350 nm which will be sufficient to facilitate the $\pi - \pi^*$ transition in the HLC molecules [36].

3. **Result and discussion**

The absorption spectrum of the HLC molecules spread onto quartz substrate is shown in Fig. 3. The major absorption peaks were seen at 220, 256 and 357 nm. The absorption peak corresponding to 357 nm is due to $\pi - \pi^*$ transition. The absorption at 357 nm can cause a trans-cis transformation of the azo-groups of the HLC molecules. The cis-transformed molecules can switch back to trans configuration on exposure to the ambient light or even in the dark state [9].

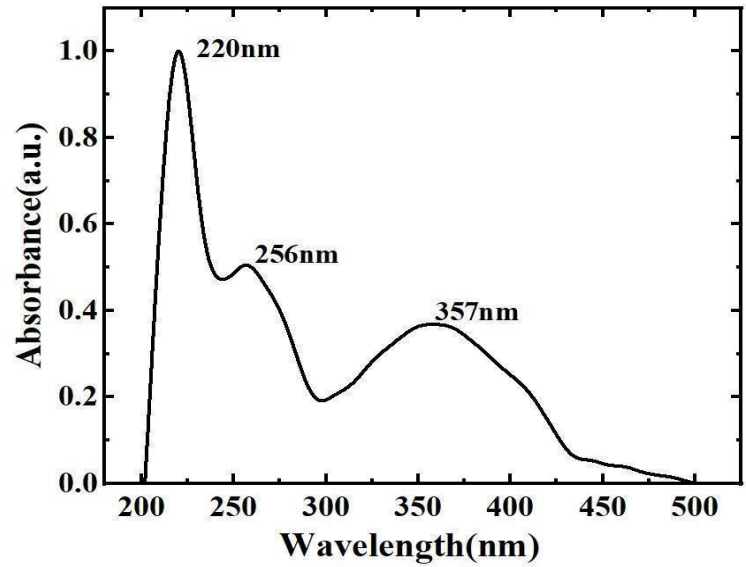

**Fig. 3.** Absorption spectrum of HLC molecules

The morphological change in the azo-based molecules due to the incidence of electromagnetic waves can be used potentially as optical switches. Similarly, for a waveguide application [37], it is essential to control the refractive index (RI) using external parameters. In this article, we report a control of RI due to photoinduced isomerization of HLC molecules in ultrathin LB film.

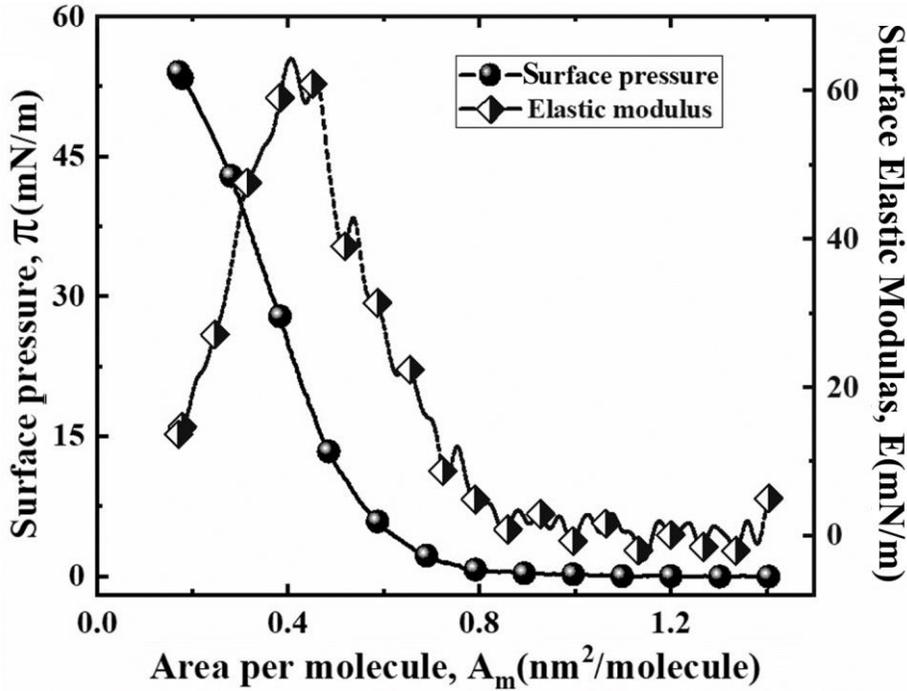

**Fig. 4.** Surface pressure (π) – Area per molecule ($A_m$) isotherm and the corresponding in-plane surface elastic modulus (E) - $A_m$ of Langmuir monolayer of HLC molecules at the air-water interface.

The π − $A_m$ isotherm of HLC molecules at the air-water interface under the dark condition is shown in Fig. 4. The dark condition was chosen to ensure that all the azo-group of the molecules should exist in trans-state before deposition of ultrathin film using LB technique. The isotherm of the monolayer of HLC molecules shows a classical trend. It shows a startup rise in surface pressure at around 0.7 nm². The surface pressure continues to rise monotonically thereafter till a change in slope is noticed at 0.3 nm². This might be the initiation of a collapsed state. In plane surface elastic modulus (E) is calculated from the π - $A_m$ isotherm using the relation E = - $A_m$ (dπ /d$A_m$) and shown in Fig. 4. A maximum value of E was found to be 65 mN/m at around 0.4 nm². This value may indicate a liquid-like phase of the HLC monolayer at the air-water interface [38]. The images of the monolayer at the air-water interface were captured using a BAM. These are shown in Fig. 5. The image captured at 1.0 nm² shows a coexistence of two features, dark and bright domains (Fig. 5(a)). The dark region represents the gas phase whereas the bright domains may represent liquid-like phase of the HLC monolayer. On further compression, the bright domains merge to yield a homogeneous bright texture (Fig. 5(b)). This is the uniform liquid-like phase of the monolayer of HLC molecules. A close observation of the BAM image of the liquid-like phase shows some patchy texture.

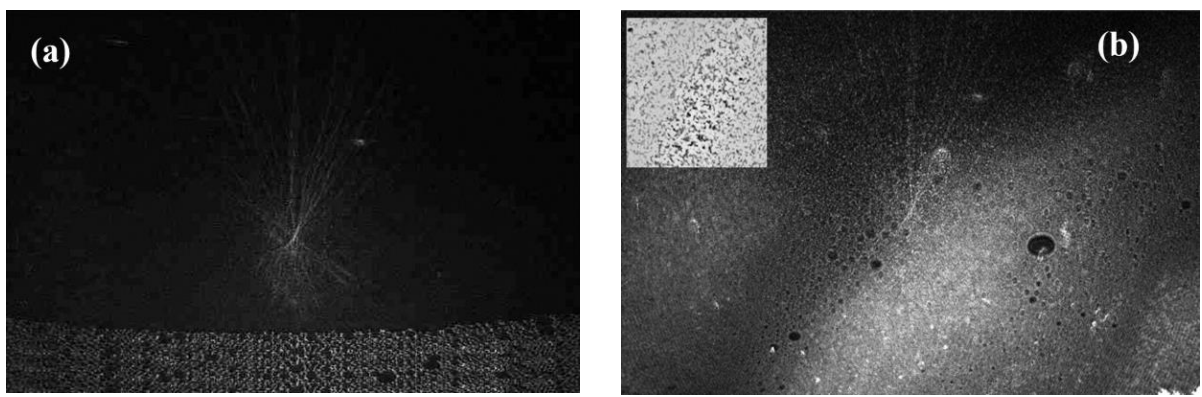

**Fig. 5.** The BAM images were taken at an area per molecule of (a) 1.0 nm$^2$ and (b) 0.5 nm$^2$. The size of the images is 1.2×1.8 mm$^2$. The size of the inset image is 275×225 μm$^2$.

The area of HLC molecule planar to the water surface (face-on) is estimated to be 4 nm$^2$ whereas vertical to the water surface (edge-on) configuration is estimated to be 0.35 nm$^2$. The limiting area per molecule ($A_o$) for the liquid-like phase was observed 0.56 nm$^2$. This is higher than the edge-on configuration and less than the face-on configuration. It is therefore possible that the HLC molecules in liquid-like phase can have edge-on conformation with tilted molecules as shown in the schematic Fig.6 (left).

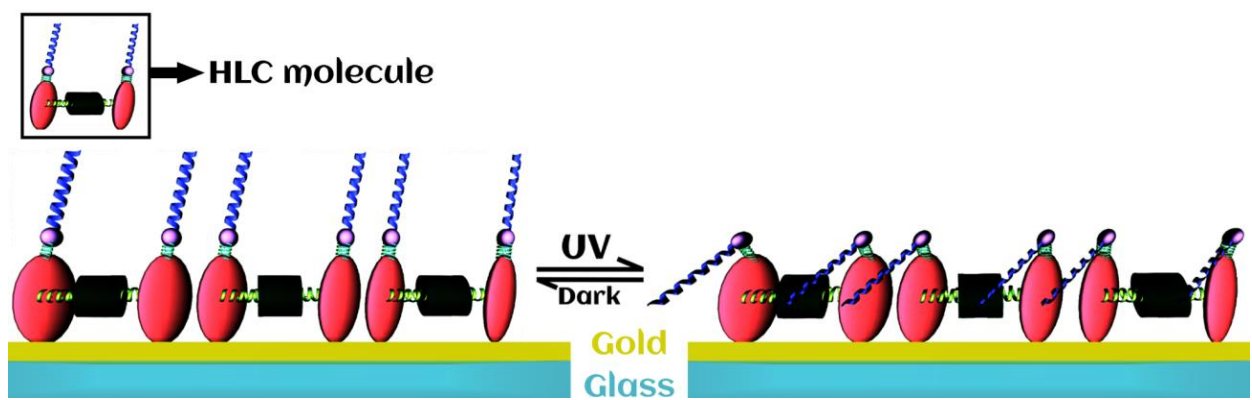

**Fig. 6.** Trans-Cis transformation of HLC molecules due to irradiation with EM wave. One HLC molecule is shown in the top-left box.

As per variation of E - $A_m$ (Fig. 4), the maximum E value was obtained at about 22 mN/m. Compression beyond this can lead to instability in the monolayer, as it approaches near to collapse. In order to achieve a stable LB film in a highly compressed and stable state, a target surface pressure of 15 mN/m was chosen for LB deposition. This pressure corresponds to the liquid-like phase of the HLC monolayer. The morphology of the LB film deposited onto silicon substrate was obtained using FESEM and shown in Fig. 7. The image shows a very interesting pattern. The dark background is due to Si wafer whereas the bright strand-like domains are due to HLC molecules in the LB film. The bright strand domains are mostly curvy in nature and they assemble to form flower-like patterns. The pattern does not exhibit any backbone which rules out the possibility of dendritic growth during natural crystallization. Since the HLC molecule exhibits several chiral centers which facilitates the strand-like domains to bend. The pattern observed in the FESEM image is due to the forced assembly of the HLC molecules under the constrained experimental conditions during LB film deposition in the liquid-like phase of the monolayer.

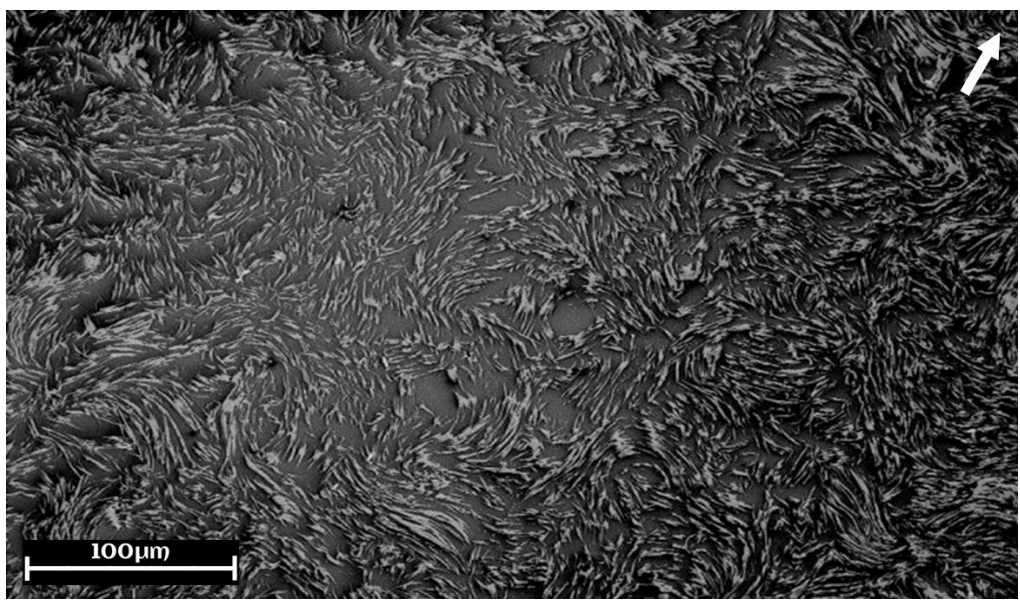

**Fig. 7.** FESEM images of LB film of HLC molecules deposited on silicon substrate at a surface pressure 15 mN/m. The arrow in the image represent dipping direction of the substrate during the LB film deposition.

The HLC molecules in the LB film can provide access to its azo-groups which can be photoinduced by the application of suitable electromagnetic waves. As observed from the absorption spectrum (Fig. 3) of the HLC molecules, an incidence of UV radiation can facilitate the trans-cis transition in the molecules which can thereby change the morphology of the molecules in the LB films. The trans-cis transformation in the molecules can be perceived by a high resolution and sensitive optical phenomenon viz. surface plasmon resonance (SPR). The morphological change in the ultrathin film can lead to changes in the refractive index (dielectrics) of the film which can be measured using the SPR phenomenon. In recent reports from our group, it has been observed that the morphological changes in aliphatic chains on organic molecules can be measured using high resolution SPR phenomenon [24,39]. In this article, trans-cis isomerization in the HLC molecules in the LB film was photoinduced by the irradiation with UV electromagnetic wave. The cis-configured molecules can switch back to trans-configuration due to irradiation with ambient light/dark state.

The LB film of HLC molecules was deposited in the liquid-like phase at a target surface pressure of 15 mN/m on the sensing chip of the SPR instrument. The SPR spectra were collected by changing the angle of incidence and recording the reflected intensity. At resonance, the reflected intensity reduces to minimum which indicates highest energy transfer from incident EM wave to surface plasmon polaritons. The angle of incidence at minimum reflected intensity is termed as resonance angle (RA). A perturbation in the thin film due to some external parameter can change the dielectrics (refractive index) of the film which in turn can shift the resonance angle. A measure of shift in resonance angle can be quantified in terms of change in refractive index using the standard Fresnel's theory of reflection from interfaces [40].

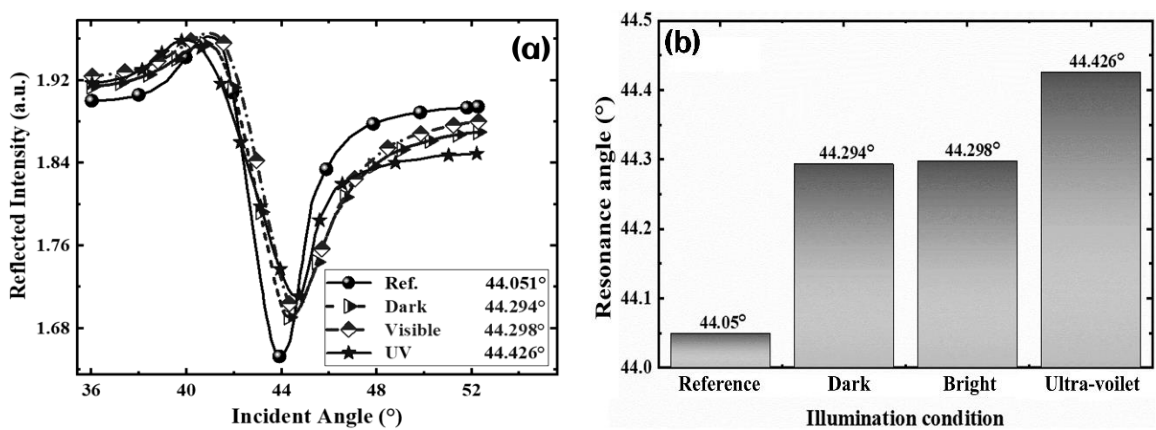

**Fig. 8.** (a) SPR spectra and (b) the corresponding resonance angle (RA) obtained from LB films of HLC molecules deposited in the liquid-like phase under the different illumination conditions. The dark and bright illumination represent the experiment performed with ambient light switch-off and switch-on, respectively.

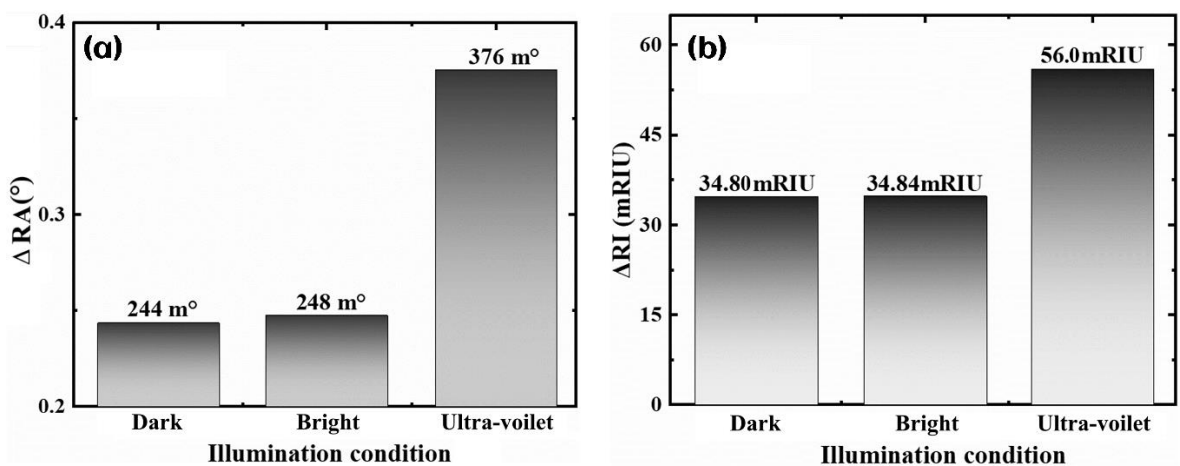

**Fig. 9.** (a) Shift in resonance angle (ΔRA) with respect to the reference (gold/air) SPR spectra and (b) the corresponding change in refractive index (ΔRI) obtained from LB films of HLC molecules deposited in the liquid-like phase under the different illumination conditions. The dark and bright illumination represent the experiment performed with ambient light switch-off and switch-on, respectively.

Fig. 8(a) shows the SPR spectra of LB film of HLC molecules under the influence of irradiation with different illumination conditions. The shift in the spectra towards higher angle of incidence as compared to that of reference (gold/air) indicates dielectric perturbation due to deposition of LB films of HLC molecules and its dependency on irradiation with EM waves. It can be noted from the bar diagram (Fig. 8(b)) that RA measured under dark and bright states are nearly the same. This may indicate that the structural perturbation is not induced either in the dark or bright state of the experimental measurements. The RA shifted to higher values due to irradiation with UV light. This is due to trans-cis transformation of the HLC molecules due to the UV light leading to change in refractive index of the film. The shift in resonance angle (ΔRA) with respect to the reference (gold/air) and the corresponding change in refractive index (ΔRI) of the LB film of HLC molecules under the different illumination condition is shown in Fig. 9. The RI was calculated using Fresnel's reflection theory for a stack of layers [33]. The change in RI due to deposition of LB film is found to be 34.8 mRIU. This value remains invariant either in the dark state or in the bright state. However, a significant increase in the value of ΔRI (56 mRIU) was found when the LB film was irradiated with UV light. This is due to trans-cis conformational change in the HLC molecules of the LB film. We have observed full recovery of the cis-state to trans-state by simply maintaining the dark condition during the SPR measurement.

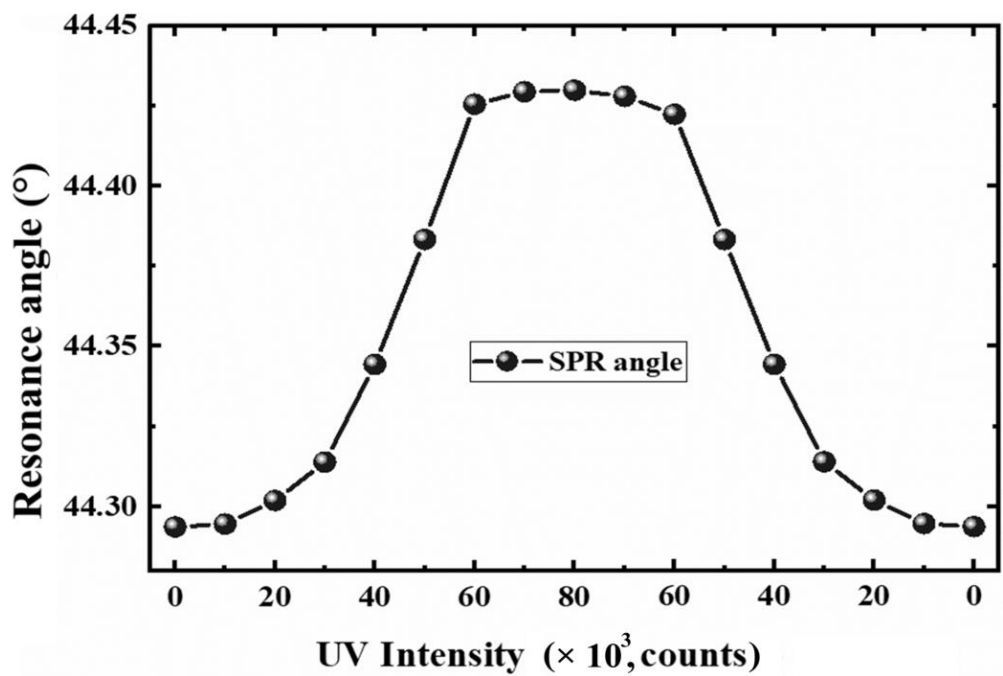

**Fig. 10.** Optical switching due to UV illumination

The morphological changes due to UV irradiation induced isomerization can find potential applications in the field of optical switches and RI modulated waveguides. The extent of isomerization was explored by increasing the intensity of the UV radiation and measuring the RA, simultaneously. This is shown in Fig. 10. It appears from the curve that there is a threshold intensity of 10000 counts of the UV radiation below which the isomerization does not take place. The RA rises monotonically thereafter till it reaches its maximum at 60000 counts. The RA saturates above 60000 counts of the incident UV radiation. This cycle repeats perfectly which indicates that there is no permanent deformation in the molecular conformation due to the UV irradiation and a complete recovery of the molecules from cis to trans-state. The UV radiation above 60000 counts can switch the number of molecules to the highest extent. These features are a good indication of the optical switch. The modulation in RA can be looked upon as the equivalent modulation of the refractive index of the LB film. Therefore, our studies suggest a precise control of the refractive index of thin film as a function of intensity of the UV radiation. The normalized rate of isomerization (NRI in %) is calculated from the slope of RA Vs intensity curve of Fig. 10 and shown in Fig. 11. The NRI can be useful for the prediction of switch-on and switch-off intensity. The intensity below 10% of NRI and above 90% of NRI can be considered as switch-off and switch-on intensity of the optical device. It can be noted from Fig. 11 that switch-on and switch-off intensity of the optical device based on LB film of HLC molecules are 13000 and 46000 counts, respectively.

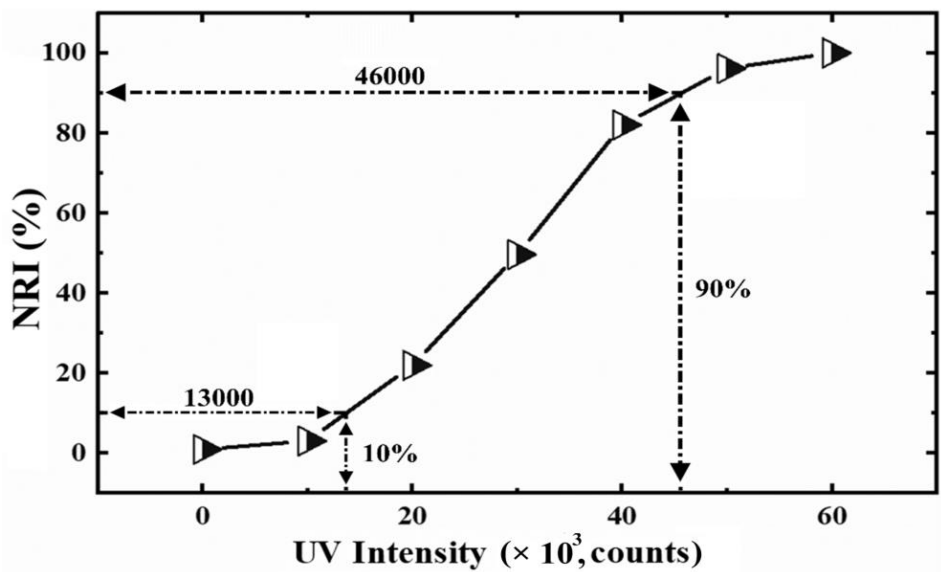

**Fig. 11.** Normalized rate of isomerization (NRI) of the HLC molecules in the LB film as a function of intensity of UV light irradiation.

## 4. Conclusion

The HLC molecules exhibiting optically switchable Azo-groups can form a very stable Langmuir monolayer at the air-water interface. The monolayer exhibited gas and liquid-like phases. A highly optically active layer was created by the deposition of LB film of the HLC molecules in the liquid-like phase. The photo-isomerization of the azo-group of the HLC molecules in the LB film due to irradiation with a UV light can lead to morphological change. Such morphological changes can lead to a miniscule change in the refractive index of the film which can be measured using a high resolution and sensitive SPR phenomenon. A large change in the refractive index (56 mRIU) of the LB film of HLC molecules on irradiation with UV light was found. Such change clearly indicates the morphological transformation due to trans-cis isomerization of the HLC molecule due to irradiation with the UV light. The study on the extent of isomerization indicates that intensity of UV source less than 13000 and more than 46000 can be considered switch-off and switch-on of the optical device developed using the LB film of HLC molecules. This study also suggests that the LB film of HLC molecule can be employed in optical waveguides wherein the local refractive index can be controlled by the suitable irradiation of UV light.

*CRediT authorship contribution statement*

**Ashutosh Joshi :** Experiments, analysis of data, manuscript drafting. **Akash Gayakwad:** Analysis of data, part of manuscript drafting. **Manjuladevi. V:** Editing and final manuscript preparation. **Mahesh Varia:** Synthesis and characterization of HLC. **S. Kumar:** Synthesis and characterization of HLC. **R K Gupta:** Conceptualization, editing and supervision and final draft of manuscript.

**Declaration of Competing Interest**

The authors declare that they have no known competing financial interests or personal relationships that could have appeared to influence the work reported in this paper.

**Acknowledgment**

The authors are thankful to SERB (CRG/2018/000755) and IDP (IDP/SEN/06/2015), DST India for their support. This is a post-peer-review, pre-copyedit version of an article published in J. Molecular Liquids (Elsevier). The final authenticated version is available online at: https://doi.org/10.1016/j.molliq.2022.120071

**References**

[1] T. Schultz, S. Ullrich, J. Quenneville, T.J. Martinez, M.Z. Zgierski, A. Stolow, Azobenzene photoisomerization: Two states and two relaxation pathways explain the violation of Kasha's rule., Femtochemistry Femtobiology Ultrafast Events Mol. Sci. 6 (2004) 45–48. https://doi.org/10.1016/B978-044451656-5/50008-6.

[2] M. K. Sinha, P. Mondal, R. Singh, Stimuli Responsive Polymeric Membranes, 2018. https://www.elsevier.com/books/stimuli-responsive-polymeric-membranes/purkait/978-0-12-813961-5.

[3] A. Natansohn, P. Rochon, Photoinduced motions in azo-containing polymers, Chemical Reviews, 102 (2002) 4139–4175. https://doi.org/10.1021/cr970155y.

[4] E. Blasco, M. Piñol, C. Berges, C. Sánchez-Somolinos, L. Oriol, Smart Polymers for Optical Data Storage, Second Edi, Elsevier Ltd., 2 (2019), 567-606. https://doi.org/10.1016/B978-0-08-102416-4.00016-8.

[5] H. Menzel, Photoisomerization in Langmuir-Blodgett-Kuhn Structures, Photoreactive Organic

Thin Films, Academic Press, (2002) 179–218. https://doi.org/10.1016/b978-012635490-4/50007-x.

[6] T. Seki, K. Ichimura, Dynamic Photocontrols of Molecular Organization and Motion of Materials by Two-Dimensionally Arranged Azobenzene Assemblies, Photoreactive Organic Thin Films, 36 (2002) 487–512. https://doi.org/10.1016/B978-012635490-4/50016-0.

[7] Q. Yu, H. Chen, Interaction of switchable biomaterials surfaces with proteins, Switchable Responsive Surfaces Materials for Biomedical Applications (2015) 167–188. https://doi.org/10.1016/B978-0-85709-713-2.00007-9.

[8] H. Rau, Photoisomerization of Azobenzenes, Photoreact. Organic Thin Films (2002) 3–47. https://doi.org/10.1016/B978-012635490-4/50002-0.

[9] S.-W. Oh, J.-M. Baek, S.-H. Kim, T.-H. Yoon, Optical and electrical switching of cholesteric liquid crystals containing azo dye, RSC Advances, 7 (2017) 19497–19501. https://doi.org/10.1039/c7ra01507k.

[10] J. Hou, A. Mondal, G. Long, L. de Haan, W. Zhao, G. Zhou, D. Liu, D.J. Broer, J. Chen, B.L. Feringa, Photo-responsive Helical Motion by Light-Driven Molecular Motors in a Liquid-Crystal Network, Angew. Chemie International Ed., 60 (2021) 8251–8257. https://doi.org/10.1002/ANIE.202016254.

[11] H. Liu, X. Liang, T. Jiang, Y. Zhang, S. Liu, X. Wang, X. Fan, X. Huai, Y. Fu, Z. Geng, D. Zhang, Analysis of structural morphological changes from 3DOM $V_2O_5$ film to $V_2O_5$ nanorods film and its application in electrochromic device, Solar Energy Materials and Solar Cells, 238 (2022) 111627. https://doi.org/10.1016/J.SOLMAT.2022.111627.

[12] S.S. Kundale, A.P. Patil, S.L. Patil, P.B. Patil, R.K. Kamat, D. kee Kim, T.G. Kim, T.D. Dongale, Effects of switching layer morphology on resistive switching behavior: A case study of electrochemically synthesized mixed-phase copper oxide memristive devices, Applied Materials Today, 27 (2022) 101460. https://doi.org/10.1016/J.APMT.2022.101460.

[13] M.M. Skolinick, Application of morphological transformations to the analysis of two-dimensional electrophoretic gels of biological materials, Computer Vision, Graphics, and Image Processing, 35 (1986) 306–332. https://doi.org/10.1016/0734-189X(86)90003-4.

[14] D. Han, X. Tong, Y. Zhao, T. Galstian, Y. Zhao, Cyclic azobenzene-containing side-chain liquid crystalline polymers: Synthesis and topological effect on mesophase transition, order, and photoinduced birefringence, Macromolecules, 43 (2010) 3664–3671. https://doi.org/10.1021/MA100246C.

[15] B.T. Hogan, E. Kovalska, M.F. Craciun, A. Baldycheva, 2D material liquid crystals for optoelectronics and photonics, Journal of Materials Chemistry C, 5 (2017) 11185–11195. https://doi.org/10.1039/C7TC02549A.

[16] J.Y. Lee, J.H. Shin, G.H. Lee, C.H. Lee, Two-Dimensional Semiconductor Optoelectronics Based on van der Waals Heterostructures, Nanomaterials, 6 (2016) 193. https://doi.org/10.3390/NANO6110193.

[17] X. Zhang, J. Shao, C. Yan, R. Qin, Z. Lu, H. Geng, T. Xu, L. Ju, A review on optoelectronic device applications of 2D transition metal carbides and nitrides, Materials & Design, 200 (2021) 109452. https://doi.org/10.1016/J.MATDES.2021.109452.

[18] Z. Gao, F. Zhang, L. Gao, X. Tian, W. Qu, T. Yang, J. Li, K. Guo, Y. Miao, An A-D-A type of thiophene derivative with morphology-determining luminescent performance: Synthesis and application in a light emitting device, Journal of Luminescence, 219 (2020) 116919. https://doi.org/10.1016/J.JLUMIN.2019.116919.

[19] P. Taneja, S.B. Khandagale, V. Manjuladevi, R.K. Gupta, D. Kumar, K.K. Gupta, Heavy Metal Ion Sensing Using Ultrathin Langmuir-Schaefer Film of Tetraphenylporphyrin Molecule, IEEE Sensors Journal, 20 (2020) 3442–3451. https://doi.org/10.1109/JSEN.2019.2959488.

[20] P. Taneja, V. Manjuladevi, R.K. Gupta, S. Kumar, K.K. Gupta, Facile ultrathin film of silver nanoparticles for bacteria sensing, Colloids and Surfaces B Biointerfaces. 196 (2020) 111335. https://doi.org/10.1016/J.COLSURFB.2020.111335.

[21] G. Roberts, ed., Langmuir-Blodgett Films, (1990). https://doi.org/10.1007/978-1-4899-3716-2.

[22] C. Fang, I. Yoon, D. Hubble, T.N. Tran, R. Kostecki, G. Liu, Recent Applications of Langmuir–Blodgett Technique in Battery Research, ACS Applied Materials & Interfaces, 14 (2022) 2431–2439. https://doi.org/10.1021/acsami.1c19064.

[23] O.N. Oliveira, L. Caseli, K. Ariga, The Past and the Future of Langmuir and Langmuir-Blodgett Films, Chemical Reviews, 122 (2022) 6459–6513. https://doi.org/10.1021/acs.chemrev.1c00754.

[24] A. Kumar, R.K. Gupta, V. Manjuladevi, A. Joshi, Surface Plasmon Resonance for In-Plane Birefringence Measurement of Anisotropic Thin Organic Film, Plasmonics. 16 (2021) 1023–1028. https://doi.org/10.1007/s11468-021-01373-1.

[25] A. Joshi, A. Kumar, V. Manjuladevi, R.K. Gupta, Modulating surface plasmon resonance response by an external electromagnetic wave, Europhysics Letters, 133 (2020). https://doi.org/10.1209/0295-5075/133/67005.

[26] A. Modlińska, K. Inglot, T. Martyński, R. Dąbrowski, J. Jadżyn, D. Bauman, A. Modlin´ska, M. Modlin´ska, T. Martyn´ski, M. Martyn´ski, R. Dabrowski B, J. Jad, Influence of the molecular structure of thermotropic liquid crystals on their ability to form monolayers at interface, Liquid Crystals, 36(2009), 197–208. https://doi.org/10.1080/02678290902759236.

[27] R.K. Gupta, V. Manjuladevi, C. Karthik, K. Choudhary, Liquid Crystals Thin films of discotic liquid crystals and their applications, Liquid Crystals, 43(2016), 2079–2091. https://doi.org/10.1080/02678292.2016.1195454.

[28] K. Choudhary, R.K. Gupta, R. Pratibha, B.K. Sadashiva, V. Manjuladevi, Alignment of liquid crystals using Langmuir–Blodgett films of unsymmetrical bent-core liquid crystals, Liquid Crystals, 46 (2019) 1494–1504. https://doi.org/10.1080/02678292.2019.1574035.

[29] A. Yoshizawa, Unconventional liquid crystal oligomers with a hierarchical structure, Journal of Materials Chemistry, 18 (2008) 2877–2889. https://doi.org/10.1039/B802712A.

[30] M. Okuno, D. Ishikawa, W. Nakanishi, K. Ariga, T.A. Ishibashi, Symmetric Raman Tensor Contributes to Chiral Vibrational Sum Frequency Generation from Binaphthyl Amphiphile Monolayers on Water: Study of Electronic Resonance Amplitude and Phase Profiles, The Journal of Physical Chemistry C 125 (2021) 25356. https://doi.org/10.1021/ACS.JPCC.1C09228.

[31] J. Holec, J. Rybáček, J. Vacek, M. Karras, L. Bednárová, M. Buděšínský, M. Slušná, P. Holý, B. Schmidt, I.G. Stará, I. Starý, Chirality-Controlled Self-Assembly of Amphiphilic

Dibenzo[6] helicenes into Langmuir–Blodgett Thin Films, Chemistry-A European Journal, 25 (2019) 11494–11502. https://doi.org/10.1002/CHEM.201901695.

[32] J.A. Zasadzinski, R. Viswanathan, L. Madsen, J. Garnaes, D.K. Schwartz, Langmuir-Blodgett films, Science (New York, N.Y.), 263(5154), 1726–1733. https://doi.org/10.1126/SCIENCE.8134836.

[33] V.P. Devanarayanan, V. Manjuladevi, R.K. Gupta, Surface plasmon resonance sensor based on a new opto-mechanical scanning mechanism, Sensors and Actuators B Chemical, 227 (2016) 643–648. https://doi.org/10.1016/J.SNB.2016.01.027.

[34] R.K. Gupta, Sensing Through Surface Plasmon Resonance Technique, (2017) 39–53. https://doi.org/10.1007/978-3-319-48081-7_3.

[35] J. Homola, Surface plasmon resonance sensors for detection of chemical and biological species, Chemical Reviews, 108 (2008) 462–493. https://doi.org/10.1021/cr068107d.

[36] O.N. Oliveira, M. Raposo, A. Dhanabalan, Langmuir-Blodgett and Self-Assembled Polymeric Films, Handbook Surfaces and Interfaces of Materials, Elsevier (2001) 1–63. https://doi.org/10.1016/B978-012513910-6/50047-5.

[37] E. Heydari, E. Mohajerani, A. Shams, All optical switching in azo-polymer planar waveguide, Optics Communications, 284 (2011) 1208–1212. https://doi.org/10.1016/J.OPTCOM.2010.10.096.

[38] R.K. Gupta, V. Manjuladevi, Liquid Crystals at Interfaces, Israel Journal of Chemistry, 52 (2012) 809–819. https://doi.org/10.1002/IJCH.201200030.

[39] V.P. Devanarayanan, V. Manjuladevi, M. Poonia, R.K. Gupta, S.K. Gupta, J. Akhtar, Measurement of optical anisotropy in ultrathin films using surface plasmon resonance, Journal of Molecular Structure, 1103 (2016) 281–285. https://doi.org/10.1016/J.MOLSTRUC.2015.09.018.

[40] K. Kurihara, K. Nakamura, K. Suzuki, Asymmetric SPR sensor response curve-fitting equation for the accurate determination of SPR resonance angle, Sensors and Actuators B: Chemical,